\documentclass[prl,twocolumn,amsmath,amssymb,10pt,aps,longbibliography,superscriptaddress,citeautoscript,bibnotes,floatfix]{revtex4-2}

\usepackage{graphicx} 
\usepackage{dcolumn}
\usepackage{bm}
\usepackage{graphicx,color,xcolor}
\usepackage{caption}
\usepackage{physics}
\usepackage{comment}
\usepackage{ulem}
\graphicspath{ {./Figures/} }
\usepackage{textcomp}
\usepackage[utf8]{inputenc}
\usepackage[T1]{fontenc}
\usepackage{amsmath}
\usepackage{amssymb}
\usepackage{amsthm}
\usepackage{siunitx}
\usepackage{physics}
\usepackage{graphicx}
\usepackage{subfigure}
\usepackage{tabularx}
\usepackage{booktabs}
\usepackage{multirow}
\usepackage[colorlinks=true,linkcolor=blue,citecolor=blue]{hyperref} 

\usepackage{xcolor}

\newcommand*{\diff}{\mathop{}\!\mathrm{d}} 

\usepackage{soul}

\begin{document}

\title{Non-Abelian Fractional Quantum Anomalous Hall States and First Landau Level Physics in Second Moiré Band of Twisted Bilayer MoTe${}_2$}

\author{Cheong-Eung Ahn}
\thanks{These authors contributed equally.}
\affiliation{Department of Physics, Pohang University of Science and Technology, Pohang, 37673, Republic of Korea}
\affiliation{Center for Artificial Low Dimensional Electronic Systems, Institute for Basic Science, Pohang 37673, Korea}

\author{Wonjun Lee}
\thanks{These authors contributed equally.}
\affiliation{Department of Physics, Pohang University of Science and Technology, Pohang, 37673, Republic of Korea}
\affiliation{Center for Artificial Low Dimensional Electronic Systems, Institute for Basic Science, Pohang 37673, Korea}

\author{Kunihiro Yananose}
\affiliation{Korea Institute for Advanced Study, Seoul 02455, Korea}

\author{Youngwook Kim}
\affiliation{Department of Physics and Chemistry, Daegu Gyeongbuk Institute of Science and Technology (DGIST), Daegu 42988, Republic of Korea}

\author{Gil Young Cho}
\thanks{gilyoungcho@postech.ac.kr}
\affiliation{Department of Physics, Pohang University of Science and Technology, Pohang, 37673, Republic of Korea}
\affiliation{Center for Artificial Low Dimensional Electronic Systems, Institute for Basic Science, Pohang 37673, Korea}
\affiliation{Asia-Pacific Center for Theoretical Physics, Pohang, Gyeongbuk, 37673, Korea}

\date{\today} 

\begin{abstract}
Utilizing the realistic continuum description of twisted {bilayer} MoTe$_2$ and many-body exact diagonalization calculation, we establish that the second moiré band of twisted {bilayer} MoTe$2$, at a small twist angle of approximately $2^\circ$, serves as an optimal platform for achieving the long-sought non-Abelian fractional quantum anomalous Hall states without the need for external magnetic fields. Across a wide parameter range, our exact diagonalization calculations reveal that the half-filled second moiré band demonstrates the ground state degeneracy {and spectral flows}, which are consistent with the pfaffian state in the first Landau level. We further elucidate that the emergence of the non-Abelian state is deeply connected to the remarkable similarity between the second moiré band and the first Landau level. Essentially, the band not only exhibits characteristics akin to the first Landau level, {$\frac{1}{2\pi}\int_\mathrm{BZ}\diff^2\mathbf{k}\:\trace\eta(\mathbf{k}) \approx 3$} {where $\eta_{ab}(\mathbf{k})$ is the Fubini-Study metric of the band}, but also that its projected Coulomb interaction closely mirrors the Haldane pseudopotentials of the first Landau level. Motivated from this observation, we introduce a novel metric of ``first Landau level''-ness of a band, which quantitatively measures the alignment of the projected Coulomb interaction with the Haldane pseudopotentials in Landau levels. This metric is then compared with the global phase diagram {of the half-filled second moire band}, revealing its utility in predicting the parameter region of the non-Abelian state. {In addition, we uncover that the first and third moiré bands closely resemble the lowest and second Landau levels, revealing a remarkable sequential equivalence between the moiré bands and Landau levels.} We finally discuss the potential implications on experiments. 
\end{abstract}


\maketitle

Twisted moiré materials have unveiled various exciting quantum phenomena, encompassing Mott insulators \cite{cao2018correlated, regan2020mott}, strange metals \cite{cao2020strange}, and superconductivity \cite{cao2018unconventional, chen2019signatures}. Even more surprisingly, recent transport experiments on twisted MoTe$_2$ (tMoTe$_2$) and pentalayer graphene systems uncovered the emergence of fractional quantum anomalous Hall states (FQAH) and composite Fermi liquids at zero external magnetic fields \cite{cai2023signatures,zeng2023thermodynamic,park2023observation,xu2023observation,lu2024fractional,kang2024observation}, which strikingly parallels the physics of quantum Hall states under large magnetic fields in the lowest Landau levels. While the discovery of the FQAH in experiments is interesting, it is noteworthy that the states observed to date align with the Abelian quantum Hall states \cite{li2021spontaneous,crepel2023anomalous,reddy2023fractional,wang2024fractional,abouelkomsan2024band,dong2023composite,xu2024maximally,yu2023fractional,reddy2023toward}, which only accommodate the Abelian anyons. On the other hand, the anomalous version of non-Abelian fractional quantum Hall states in the first Landau level, which could lay the practical groundwork for the fault-tolerant topological quantum computer and thus have immense importance \cite{nayak2008non}, remains undiscovered. 

In this Letter, we demonstrate that the \textit{second} moiré band of tMoTe$_2$ around the twist angle $\theta \sim 2^\circ$ is an ideal platform for the realization of this long-sought non-Abelian FQAHs. This is in a sharp contrast to the previous research, which has focused on the \textit{first} moiré band of tMoTe$_2$ at \textit{larger twist angles} $\theta \sim 4^{\circ}$ \cite{dong2023composite,wang2024fractional,abouelkomsan2024band,xu2024maximally,yu2023fractional,reddy2023toward}. Our analysis is based on the continuum description of the density functional theory (DFT) band structure \cite{zhang2023polarization} and many-body exact diagonalization calculation (ED). Our ED calculation finds that when half-filled, the second moiré band gives rise to the ground state degeneracy, which is consistent with the putative non-Abelian fractional quantum Hall states in the first Landau level, i.e., pfaffian state~\cite{read2000paired}. Notably, tMoTe$_2$ and twisted bilayer WSe$_2$ with the small twist angles are readily available in experiments \cite{kang2024observation, kang2024observation2}, which calls for further experiments. 

Interestingly, we find that behind the emergence of the non-Abelian FQAHs, the striking resemblance of of the second moiré band and the first Landau level (1LL) exists. This contrasts with the first moiré band of tMoTe$_2$ for $\theta \gtrsim 3.9^\circ$, which is widely recognized for its resemblance to the lowest Landau Level (LLL) \cite{devakul2021magic,xu2024maximally,dong2023composite,reddy2023toward,morales2024magic,yu2023fractional,jia2023moire,crepel2023chiral}. Essentially, we observe that the projected Coulomb interaction closely mirrors the Haldane pseudo-potential in 1LL while the band fulfills the known 1LL trace condition {$\frac{1}{2\pi}\int_\mathrm{BZ}\diff^2\mathbf{k}\:\trace\eta(\mathbf{k}) \approx 3$}, {where $\eta_{ab}(\mathbf{k}) = \Re\left\{\mel{\partial_{k_a}u(\mathbf{k})}{\left(\mathbb{I}-\ket{u(\mathbf{k})}\bra{u(\mathbf{k})}\right)}{\partial_{k_b}u(\mathbf{k})}\right\}$ is the Fubini-Study metric of the Bloch function $\ket{u(\mathbf{k})}$} \cite{ozawa2021relations,fujimoto2024higher}. These two conditions are in principle independent to each other, however should both be satisfied to best mimic the physics of 1LL. Motivated by this observation, we present a new metric Eq.\eqref{PotentialMeasure} for 1LL-ness of the band, which quantifies the similarity of the projected Coulomb interaction to the Haldane pseudopotentials in 1LL. Notably, the full characterization of the 1LL-like Chern bands is currently a subject of active research \cite{ozawa2021relations,fujimoto2024higher}. Our metric could offer a valuable and quantitative tool for this purpose, with wide applicability. Finally, we also compare the global phase diagram around the FQAH and the values of our new metric, revealing its utility in predicting the parameter region of the 1LL-ness.

Before proceeding further, let us compare our findings with the existing studies. First, \cite{zhang2024moore} explored the first moiré band of tMoTe$_2$ at $\theta\sim 3.9^\circ$ with three-body interactions. On the other hand, \cite{reddy2024non} investigated a skyrmion Chern band model, which {captures the essence of twisted bilayer transition metal dichalcogenides}. Within this model, \cite{reddy2024non} also pointed out the 1LL-like nature of the second moiré band and emergence of the non-Abelian FQAH, using the wavefunction overlap of the single-hole states in their model with corresponding 1LL states. Our work distinguishes itself by focusing directly on the second moiré band of a realistic continuum model near $\theta\sim 2^\circ$ and revealing 1LL-ness through the projected Coulomb interactions.



\begin{table}[t!]
\begin{tabular}{c|ccccc|c}
\hline
angle & $v_1$ & $\psi$ & $\gamma_1$ & $v_2$ & $\gamma_2$ & Ref.\\
\hline\hline
$\gg 1^\circ$ &  8    & $-89.6^\circ$   &  -8.5  &   0    &  0    & \cite{wu2019topological}\\
\hline
$4.4^\circ$   & 11.2  &  $-91^\circ$    & -13.3  &   0    &  0    & \cite{reddy2023fractional}\\
\hline
$3.9^\circ$   &  7.5  & $-100^\circ$    & -11.3  &   0    &  0    & \cite{reddy2023fractional}\\
-             &  9.2  &  $-99^\circ$    & -11.2  &   0    &  0    & \cite{xu2024maximally}\\
-             & 20.8  & $-107.7^\circ$  & -23.8  &   0    &  0    & \cite{wang2024fractional}\\
-             & 17.0  & $-107.7^\circ$  & -16.0  &   0    &  0    & \cite{wang2023topology}\\
-             & 16.5  & $-105.9^\circ$  & -18.8  &   0    &  0    & \cite{jia2023moire}\\
-             &  7.94 &  $-88.43^\circ$ & -10.77 &  20.00 & 10.21 & \cite{jia2023moire}\\
-             &  9.45 &  $-85.23^\circ$ & -12.20 &  24.99 & 13.12 & \cite{zhang2023polarization,supp}\\
\hline
$2.1^\circ$   & 20.51 &  $-61.49^\circ$ &  -7.01 &  -9.08 & 11.08 & \cite{zhang2023polarization,supp}\\
\hline
\end{tabular}
\caption{\textbf{Continuum model parameters.} The parameters $v_1$, $\gamma_1$, $v_2$, $\gamma_2$ are given in units of meV.}
\label{tab:continuum_model_parameters}
\end{table}

\begin{figure}[t!]
\centering
\includegraphics[width=0.99\linewidth]{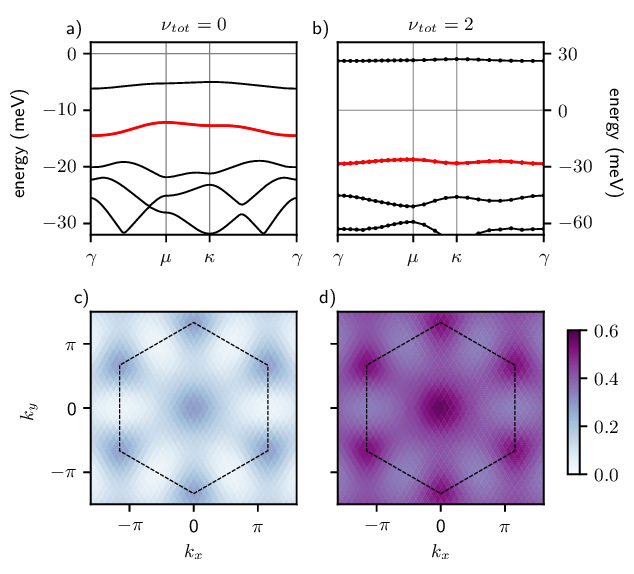}
\caption{(a) Band structure at $\nu_h=0$ and (b) its SCHF-corrected band structure at $\nu_h=2$ with $\epsilon_r=5.0$. (c) Berry curvature and (d) trace of the Fubini-Study metric of the SCHF-corrected second moiré band. The black dashed line represents the moiré BZ with $a_M=1$.}
\label{fig:MFig_1}
\end{figure}

\textbf{1. Continuum Model.} We start with the continuum model \cite{wu2019topological} of tMoTe$_2$ around the $K$-valley (with $\hbar=1$) 
\begin{equation}
\hat{h}_\uparrow=\begin{bmatrix}
-\frac{(\hat{\mathbf{k}}-K_+)^2}{2m^*}+V_+(\hat{\mathbf{r}}) & \Gamma^*(\hat{\mathbf{r}})\\
\Gamma(\hat{\mathbf{r}}) & -\frac{(\hat{\mathbf{k}}-K_-)^2}{2m^*}+V_-(\hat{\mathbf{r}})
\end{bmatrix},\label{cont}
\end{equation}
where $m^*$ is the effective mass of the monolayer electron, $K_\pm=R_{\pm\theta/2}K$ are the twisted $K$ points of the top and bottom layers, $V_\pm(\mathbf{r})$ and $\Gamma(\mathbf{r})$ are the intralayer and interlayer moiré potentials, respectively. The Hamiltonian around the $(-K)$-valley is the time-reversal conjugate of Eq.\eqref{cont}. There have been several DFT studies on tMoTe$_2$ at various twist angles~\cite{reddy2023fractional,xu2024maximally,wang2024fractional,wang2023topology,jia2023moire,zhang2023polarization,mao2023lattice}. For our focus, i.e., tMoTe$_2$ near $\theta\sim 2.1^{\circ}$, we find the parameters ($v_1,\psi,\gamma_1,v_2,\gamma_2$) of the moiré potentials by fitting the band structure of Eq.\eqref{cont} to that in \cite{zhang2023polarization} along the high-symmetry lines in momentum space~\cite{supp}, up to valley-reduced inversion symmetric second harmonics~\cite{jia2023moire,mao2023lattice,supp}. The resulting parameters are listed in Table \ref{tab:continuum_model_parameters}, whose band structure is in [Fig.\ref{fig:MFig_1}(a)].

A few comments are in order. First, we confirmed that the model parameters for $\theta \sim 2^{\circ}$ produces the series of the valley Chern numbers, consistent with the DFT calculation \cite{zhang2023polarization} and also the experiment \cite{kang2024observation}. This supports the validity of our continuum model Eq.\eqref{cont} beyond the numerical fitting. {Specifically}, each of the three moiré bands ($a=1,2,3$) exhibit a valley-projected Chern number of $C_{\uparrow, a} = +1$ with $C_{\uparrow, a} = - C_{\downarrow, a}$. Secondly, from Table \ref{tab:continuum_model_parameters}, we observe a significant renormalization of the model parameters for $\theta\sim 2.1^\circ$ relative to those for $\theta\gtrsim 3.9^\circ$ \cite{wu2019topological,reddy2023fractional,xu2024maximally,wang2024fractional,jia2023moire,wang2023topology,zhang2023polarization}. As discussed in \cite{zhang2023polarization,mao2023lattice,jia2023moire}, the higher harmonics in the model parameters are expected to play a significant role at smaller twist angles, and they can even undergo a rapid change in sign with twist angle. This is reflected in the sign of $v_2$ for $\theta\sim 2.1^\circ$ relative to those for $\theta\gtrsim 3.9^\circ$. Despite of these drastic differences in model parameters, the moiré band widths at $\theta\sim 2.1^\circ$ remain small, approximately 2.4 meV, signifying the importance of the interactions.




\textbf{2. Coulomb Interaction.} We assume the Coulomb interaction~\cite{li2021spontaneous,crepel2023anomalous,reddy2023fractional,wang2024fractional,abouelkomsan2024band,dong2023composite,xu2024maximally,yu2023fractional,reddy2023toward} 
\begin{equation}
H=-\hat{h}+\frac{1}{2A}\sum_{\mathbf{p}}V_{\mathbf{p}}{:\hat{\rho}_{\mathbf{p}}\hat{\rho}_{-\mathbf{p}}:}. \nonumber
\end{equation}
Here, $\hat{\rho}_{\mathbf{p}}=\sum_{\tau,\mathbf{k},\ell}\hat{c}_{\tau,\mathbf{k}+\mathbf{p},\ell}^\dagger\hat{c}_{\tau,\mathbf{k},\ell}$ is the density operator, $A$ is the sample area, and the normal ordering is against $\nu_{h}=0$, and  
\begin{equation}
V_{\mathbf{p}}=\left(\frac{e^2}{4\pi\epsilon_0}\right)\left(2\pi\epsilon_r^{-1}\right)\frac{\tanh(p\xi/2)}{p}, \nonumber
\end{equation}
where $\xi$ is the gate distance and $\epsilon_r$ is the relative permittivity. For simplicity, we set $\xi=\infty$, i.e., the unscreened Coulomb interaction, as its precise value has little effect on FQAHs due to the large size of the moiré unitcell \cite{yu2023fractional}.

As the DFT band is obtained at $\nu_h=0$, we use the self-consistent Hartree-Fock (SCHF) method to obtain the renormalized second moiré band at $\nu_h=2$, its Berry curvature and trace of the Fubini-Study metric [Fig.\ref{fig:MFig_1}(b-d)]. The Hartree-Fock correction makes the Berry curvature and trace of the Fubini-Study metric significantly more uniform, similar to the valley-polarized first moiré band at $\nu_h=1$~\cite{dong2023composite,reddy2023fractional,yu2023fractional}. {Some details on SCHF can be found in \cite{supp}.} On the other hand, the trace of the Fubini-Study metric, i.e., $\trace\eta(\mathbf{k})$, is much larger than the absolute Berry curvature, suggesting that the second moiré band is distinct from LLL-like bands \cite{parameswaran2012fractional,parameswaran2013fractional,ledwith2023vortexability}. Indeed, the quantum weight {$\frac{1}{2\pi}\int_\mathrm{BZ}\diff^2\mathbf{k}\:\trace\eta(\mathbf{k})$} for $\epsilon_r=5$ reaches $3.04$ approximately, well satisfying a property of 1LL-like bands~\cite{ozawa2021relations,fujimoto2024higher}, {whereas the Chern number is $C=1$}. We note that while a mathematically precise criteria for 1LL-like bands has recently been proposed in \cite{fujimoto2024higher}, it requires the global information of band structures. Hence, a direct characterization of 1LL-like bands from local {quantum geometry} has yet to be found, as noted in \cite{fujimoto2024higher}.

\begin{figure}[t!]
\centering
\includegraphics[width=0.99\linewidth]{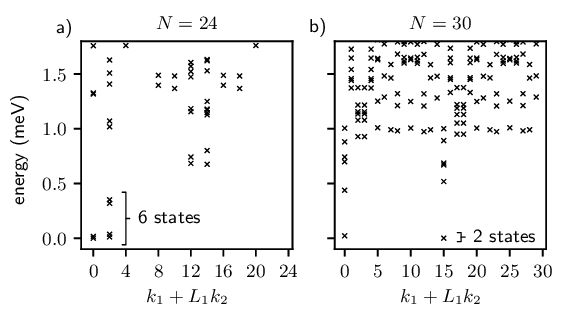}
\caption{Energy spectrum on a torus of (a) $4 \times 6$, and (b) $5 \times 6$ unitcells.}
\label{fig:MFig_2}
\end{figure}

\textbf{3. Non-Abelian FQAH.} Driven by the resemblance of the second moiré bands to 1LL, we undertake ED calculations to search for a non-Abelian FQAH state. Our analysis focuses on {a} half-filled, SCHF-corrected second moiré band. We first find that, for a system of $3\times 4$ unitcells, the ground state is fully valley-polarized for a wide range of twist angles $\theta\in[1.5^\circ,3.0^\circ]$ and dielectric constants $\epsilon_r\in[2,10]$ under the assumption that all model parameters in Eq.\eqref{cont} remain unchanged except $\theta$ \cite{supp}. This robust valley polarization mirrors that observed in the first moiré band within the Coulomb and dual-gate screened Coulomb interaction~\cite{li2021spontaneous,reddy2023fractional,yu2023fractional,wang2024fractional,potasz2024itinerant}. 
Building on this, we {next} narrow our ED investigation to a single valley. 



In [Fig.\ref{fig:MFig_2}(a,b)], we present the many-body energy spectra of a single valley for $4\times 6$ and $5\times 6$ unitcells at $\epsilon_r = 5.0$ and $\theta = 2.1^{\circ}$. Importantly, we observe the emergence of six-fold and two-fold degenerate ground states {for even and odd number of holes}, matching precisely those of the pfaffian state \cite{read2000paired,oshikawa2007topological}, which strongly suggests the emergence of a non-Abelian FQAH. {In \cite{supp}, we report the spectral flow and additional energy spectrum, which further support the emergence of the pfaffian state.}


\begin{figure}[t!]
\centering
\includegraphics[width=0.99\linewidth]{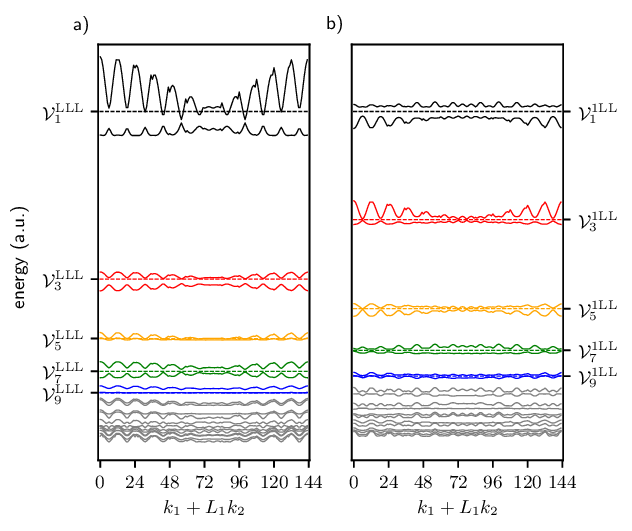}
\caption{Effective Haldane pseudopotentials of projected Coulomb interactions. The solid lines are the estimated Haldane pseudopotential of (a) the first moiré band at $\theta=3.7^\circ$ and (b) the second moiré band at $\theta=2.1^\circ$ {on a torus of $12\times 12$ unitcells}. In both cases, we take $\epsilon_r=5$. The dashed lines are the exact Haldane pseudopotentials of (a) LLL and (b) 1LL.}
\label{fig:MFig_3}
\end{figure}

\textbf{4. 1LL-like Projected Interactions.} While FQAHs require both the appropriate band structure and interactions, much of the previous literature \cite{devakul2021magic,dong2023composite,morales2024magic,yu2023fractional,jia2023moire,reddy2023fractional,reddy2024non,wang2021exact,wang2022hierarchy,wang2023origin} focuses largely on the band characteristics like the Fubini-Study metric to discuss the $n$LL-ness. Here, we focus on the nature of the projected Coulomb interaction in moiré bands and compare them with those in LLs.

To have a fair comparison of the projected Coulomb interaction in the second moiré band with that in 1LL, we calculate the effective Haldane pseudopoentials of the projected interaction from the energy spectrum of two holes~\cite{lauchli2013hierarchy}. When the band is completely flat and the projected interaction respects the translation symmetry, the energy levels of the two holes are expected to organize into pairs, irrespective of momentum. Then, the first highest pair's energy is $\mathcal{V}_1$, the second pair's energy is $\mathcal{V}_3$, the third pair's is $\mathcal{V}_5$, and so forth. If the band under consideration perfectly mimics the physics of {the} $n$LL, one expects that the pseudopotential obtained in this way will be equal to those of the {$n$}LL projected Coulomb interaction, namely the Haldane pseudopotentials $\mathcal{V}_\ell^{\text{$n$LL}}$ ($\ell=1,3,5,\cdots$), up to a global scale. For non-ideal bands, the pairs are split and show momentum dependence.

Spectacularly, we observe that the effective Haldane pseudopotentials of the projected Coulomb interaction in the SCHF-renormalized second moiré band at $\theta\sim 2.1^\circ$ [Fig.\ref{fig:MFig_1}(b)] is close to those of the 1LL-projected Coulomb interaction [Fig.\ref{fig:MFig_3}(b)]. This provides further support of the emergence of the 1LL-like non-Abelian FQAH in our ED calculation. Following the same calculation, we can also demonstrate that for the first moiré band at $\theta=3.7^\circ$, the effective pseudopotential of the projected Coulomb interaction is similar to the Haldane pseudopotential in LLL [Fig.\ref{fig:MFig_3}(a)], consistent with previous characterization of the first moiré band in terms of quantum geometry \cite{devakul2021magic,dong2023composite,morales2024magic,yu2023fractional,jia2023moire,reddy2023fractional}. For the latter, we used the continuum model Eq.\eqref{cont} with model parameters for $\theta\sim 3.9^\circ$ \cite{zhang2023polarization,supp}, also reported in Table.\ref{tab:continuum_model_parameters}. 

{By investigating the projected interactions and quantum weight, we uncover that the other moiré bands at $\theta\sim 2.1^\circ$ also closely resembles the physics of the corresponding Landau levels \cite{supp} as like the second moiré band resembles 1LL. More precisely, the effective Haldane pseudopotentials and the quantum weight $\frac{1}{2\pi}\int_\mathrm{BZ}\diff^2\mathbf{k}\:\trace\eta(\mathbf{k})$ of the third moiré band are close to those of 2LL, and those of the first moiré band are close to those of LLL. Detailed calculations and consequences of the above can be found in \cite{supp}. Hence, in tMoTe${}_2$ at $\theta\sim 2.1^\circ$, we remarkably find that the physics of the moiré bands sequentially matches those of the Landau levels.}

Motivated by the similarity of the projected interaction to the Haldane pseudopotentials, we propose a quantitative metric for the $n$LL-ness 
\begin{align}
S^2&=\frac{1}{N}\sum_{\mathbf{K}}\sum_{m=1}^\infty w_{2m-1}\left(\left|\frac{\tilde{E}_{\mathbf{K},2m-1}-E_0}{E_1-E_0}-\frac{v_{2m-1}^{\text{$n$LL}}}{v_1^{\text{$n$LL}}}\right|^2\right.\nonumber\\
&\qquad\qquad\left.+\left|\frac{\tilde{E}_{\mathbf{K},2m}-E_0}{E_1-E_0}-\frac{v_{2m-1}^{\text{$n$LL}}}{v_1^{\text{$n$LL}}}\right|^2\right) \geq 0,\label{PotentialMeasure}
\end{align} 
where $\tilde{E}_{\mathbf{K},a}$ ($a=1,2,3,\cdots$) are the energy levels of the total momentum sector $\mathbf{K}=\mathbf{k}_1+\mathbf{k}_2$ of the two holes in \textit{descending} order, and $E_1,E_0$ are the global scaling factors. Each $\ell$-th term in the sum is weighted by $w_\ell\in[0,1]$. {To correctly capture the similarity to the Landau levels, we need to set the first few weights large while $w_\ell\rightarrow 0$ for large enough $\ell$. A more detailed discussion can be found in \cite{supp}.} This metric measures both the deviation of the estimated pseudopotentials from those of $n$LL, and also the non-ideality of the band like the non-flatness and non-uniform quantum geometry. The smaller the metric $S$ is, the projected interaction to the band is closer to the Haldane pseudopotential in $n$LL. Below, we will compare its value with the many-body phase diagram.

We leave a few important remarks on Eq.\eqref{PotentialMeasure}. First, the computation of $S$ is much less costly than ED, because it requires the energy spectrum for just two particles. When $N$ is the number of unitcells, the cost is $\sim N^3$. Secondly, given that the projected Coulomb interaction does depend on the band characteristics, the comparison of the effective Haldane pseudopotential with those of the $n$LL can provide a useful alternative metric for the $n$LL-ness of Chern bands.

\begin{figure}[t!]
\centering
\includegraphics[width=0.99\linewidth]{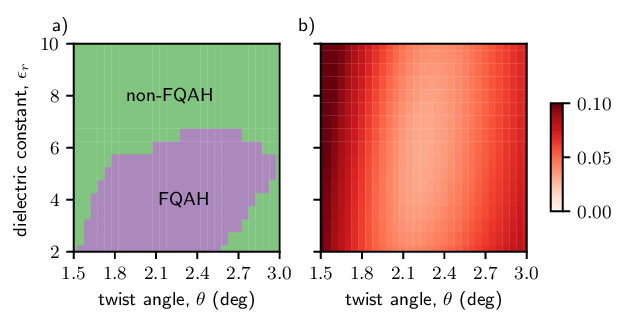}
\caption{(a) Phase diagram. (b) Value of 1LL-ness $S$, Eq.\eqref{PotentialMeasure} for $w_\ell=1$ for $\ell\leq 9$ and $w_\ell=0$ for $\ell>9$.}
\label{fig:MFig_4}
\end{figure}

\textbf{5. Phase Diagram.} Within ED, we draw the many-body phase diagram over the parameter space of $\theta\in[1.5^\circ,3.0^\circ]$ and $\epsilon_r\in[2,10]$, while holding other model parameters in Eq.\eqref{cont} invariant. Over this parameter space, we have confirmed that the system is fully valley polarized \cite{supp}, and thus we perform ED on a single valley for $N=4\times 6$ unitcells. Our primary focus is to identify the non-Abelian FQAH, whose lowest energy levels are in one-to-one correspondence with those in 1LL~\cite{reddy2024non}. The resulting phase diagram is in [Fig.\ref{fig:MFig_4}(a)], which is compared with the value of $S$ Eq.\eqref{PotentialMeasure}. We observe that the projected interaction in the second moiré band remains close to the projected Coulomb interaction in 1LL for a wide range of angles and interaction strengths, with an optimal value in the vicinity of $\theta\sim 2.2^\circ$ and $\epsilon_r\sim 6$ [Fig.\ref{fig:MFig_4}(b)]. Globally, we find that the FQAH state is stabilized when the interaction strength is large enough, and when the projected interaction is 1LL-like, i.e., $S$ is small enough, revealing the efficacy of our metric Eq.\eqref{PotentialMeasure}.

\textbf{6. Conclusions.} We have shown that the behavior of the second moiré band in tMoTe$_2$ around $\theta \sim 2.1^{\circ}$ exhibits remarkable parallels with 1LL. This similarity is not just limited to the band's characteristics {$\frac{1}{2\pi}\int_\mathrm{BZ}\diff^2\mathbf{k}\:\trace\eta(\mathbf{k}) \approx 3$}, but also extends to the nature of the projected Coulomb interaction, which closely resembles the Haldane pseudopotentials in 1LL. These observations collectively hint at the possible emergence of a non-Abelian FQAH at $\nu_h = 2+1/2$, which we have confirmed through ED calculations. Moreover, we introduced a novel metric Eq.\eqref{PotentialMeasure} of 1LL-ness of a band, which measures the similarity between the projected Coulomb interaction and the Haldane pseudopotentials in 1LL. This new metric $S$ agrees reasonably well with the region where the FQAH is stabilized [Fig.\ref{fig:MFig_4}], when the interaction is strong enough.

Finally, our theory here clearly demonstrates that the tMoTe$_2$ or twisted bilayer WSe$_2$ with small twist angles ~\cite{kang2024observation, kang2024observation2} represent a promising path for the 1LL-like non-Abelian FQAHs. In particular, in these two-dimensional moire materials, there are various tuning parameters like displacement field, interlayer distance via pressure, and substrate variations, which may lead to the eagerly-anticipated discovery of the 1LL-like non-Abelian FQAHs. \newline

{\textit{- Note Added}: Near the completion of this work, we became aware of the two similar works \cite{reddy2024non,xu2024multiple}. An updated version (v2) of \cite{reddy2024non} identified a non-Abelian FQAH in the second moiré band of a skyrmion Chern band model, which is adiabatically-connected to the continuum model of twisted bilayer MoTe${}_2$ at $\theta\sim 2.5^\circ$, while \cite{xu2024multiple} identified a non-Abelian FQAH in the second moiré band of the continuum model over a wide range of twist angles, using the parameters fitted to their DFT calculations of twisted bilayer MoTe${}_2$ at $\theta\sim 1.9^\circ$.} \newline


\acknowledgments
We thank Yong Baek Kim and Kwon Park for helpful discussions and Young-Woo Son for collaborations in related works. C.-E. A., W. L. and G.Y.C. are supported by Samsung Science and Technology Foundation under Project Number SSTF-BA2002-05 and {SSTF-BA2401-03}, the NRF of Korea (Grant No.RS-2023-00208291, No.2023M3K5A1094810, No.2023M3K5A1094813) funded by the Korean Government (MSIT),  the Air Force Office of Scientific Research under Award No.FA2386-22-1-4061, and Institute of Basic Science under project code IBS-R014-D1. K.~Y. was supported by a KIAS individual grant (CG092501) at Korea Institute for Advanced Study. A part of computations were supported by Center for Advanced Computation of KIAS. {Y. K. is supported by NRF of Korea (2020R1C1C1006914, 2022M3H3A1098408)  funded by the Korean Government (MSIT).}

\bibliographystyle{apsrev4-1}
\bibliography{refs.bib}


\end{document}


\title{Supplementary Information for "Non-Abelian Fractional Quantum Anomalous Hall States and First Landau Level Physics in Second Moiré Band of Twisted Bilayer MoTe${}_2$"}

\author{Cheong-Eung Ahn}
\thanks{These authors contributed equally.}
\affiliation{Department of Physics, Pohang University of Science and Technology, Pohang, 37673, Republic of Korea}
\affiliation{Center for Artificial Low Dimensional Electronic Systems, Institute for Basic Science, Pohang 37673, Korea}

\author{Wonjun Lee}
\thanks{These authors contributed equally.}
\affiliation{Department of Physics, Pohang University of Science and Technology, Pohang, 37673, Republic of Korea}
\affiliation{Center for Artificial Low Dimensional Electronic Systems, Institute for Basic Science, Pohang 37673, Korea}

\author{Kunihiro Yananose}
\affiliation{Korea Institute for Advanced Study, Seoul 02455, Korea}

\author{Youngwook Kim}
\affiliation{Department of Physics and Chemistry, Daegu Gyeongbuk Institute of Science and Technology (DGIST), Daegu 42988, Republic of Korea}

\author{Gil Young Cho}
\thanks{gilyoungcho@postech.ac.kr}
\affiliation{Department of Physics, Pohang University of Science and Technology, Pohang, 37673, Republic of Korea}
\affiliation{Center for Artificial Low Dimensional Electronic Systems, Institute for Basic Science, Pohang 37673, Korea}
\affiliation{Asia-Pacific Center for Theoretical Physics, Pohang, Gyeongbuk, 37673, Korea}

\setcounter{tocdepth}{0}

\maketitle
\tableofcontents

\section{Continuum Model}
Here we provide additional details on the continuum model used in the main text, Eq.(1), the fitting to DFT calculations, and the quantum geometry of the second moir\'{e} band at charge neutrality.

\paragraph{Valley-Reduced Inversion Symmetric Second Harmonics} The single-electron Hamiltonian of twisted bilayer MoTe${}_2$ systems satisfies three-fold rotational symmetry $C_3$, time-reversal symmetry $\mathcal{T}$, and two-fold rotational symmetry along an in-plane axis $C_2^y$. As discussed in \cite{jia2023moire}, when the moir\'{e} potentials are considered only up to the first harmonics, imposing the three symmetries results in an additional valley-reduced inversion symmetry $\tilde{\mathcal{I}}$,
\begin{equation}
\tilde{\mathcal{I}}c_{\tau,\ell,\mathbf{r}}\tilde{\mathcal{I}}^{-1}=c_{\tau,\overline{\ell},-\mathbf{r}},
\end{equation}
which is not a combination of $C_3,\mathcal{T},C_2^y$. Here $\tau\in\{\uparrow,\downarrow\}$ is the spin/valley index, $\ell\in\{0,1\}$ is the layer index for top and bottom layers, $\overline{\ell}=1-\ell$ denotes the opposite layer index, $\mathbf{r}$ is the atom site, and $c_{\tau,\ell,\mathbf{r}}$ is the corresponding electron annihilation operator. For physical inversion symmetry, if {it} exists, the symmetry should also reverse $\tau$, but for $\tilde{\mathcal{I}}$, $\tau$ remains invariant. This \textit{symmetry} appears because the moir\'{e} potentials are restricted to the first harmonics, and the inclusion of the higher harmonics leads to terms which respect only the $C_3,\mathcal{T},C_2^y$ symmetries while breaking this inversion \textit{symmetry} explicitly. In the DFT bands, the split of energy bands along the $\overline{\gamma\mu}$-line \cite{reddy2023fractional,xu2024maximally,wang2024fractional,wang2023topology,jia2023moire,zhang2023polarization,mao2023lattice} indicates that $\tilde{\mathcal{I}}$ is indeed broken in realistic band structures.

Imposing $C_3,\mathcal{T},C_2^y$ symmetries on the moir\'{e} potentials up to the second harmonics, we obtain the general expression
\begin{align}
V_\pm(\mathbf{r})&=2v_1\sum_{j=0}^2\cos\left(\mathbf{q}_j^M\cdot\mathbf{r}\pm\psi\right)\nonumber\\
&\qquad+2v_2\sum_{j=0}^2\cos\left(\mathbf{q}_j^{M,2}\cdot\mathbf{r}+\psi'\right),\\
\Gamma(\mathbf{r})&=\gamma_1\left(1+e^{+i\mathbf{q}_2^M\cdot\mathbf{r}}+e^{-i\mathbf{q}_1^M\cdot\mathbf{r}}\right)\nonumber\\
&\qquad+\gamma_2\left[e^{-i\mathbf{q}_0^{M,2}\cdot\mathbf{r}}+2\cos\left(\mathbf{q}_0^M\cdot\mathbf{r}\right)\right].
\end{align}
Here $\mathbf{q}_0^M=\left(4\pi/a_M\sqrt{3}\right)\hat{\mathbf{x}}$ and $\mathbf{q}_j^M=R_{(2\pi/3)j}\mathbf{q}_0^M$ are the moir\'{e} reciprocal lattice vectors, $\mathbf{q}_j^{M,2}=\mathbf{q}_{j+1}^M-\mathbf{q}_{j+2}^M$ are the next-nearest neighbor vectors of the moir\'{e} reciprocal lattice, and $a_M=a_0\left(2\sin\frac{\theta}{2}\right)^{-1}$ is the moir\'{e} lattice periodicity. Up to the second harmonics, the breaking of $\tilde{\mathcal{I}}$ is fully given by $\psi'$, i.e. when $\psi'=0$, the moir\'{e} potentials are valley-reduced inversion symmetric. As the scale of valley-reduced inversion symmetry breaking is not significant, we make the approximation $\psi'=0$, similar to \cite{jia2023moire}.

\paragraph{Fitting to DFT Calculations}
To fit the continuum model to DFT calculations, we define the objective function as the root-mean-square-error (RMSE) between the target DFT bands and continuum model bands along {the} high-symmetry crystal momentum path, {$\gamma$-$\mu$-$\kappa$-$\gamma$}. As we are mainly interested in the second band, we have weighted the RMSE so the contribution of the third band is half of that of the first and second bands, and we ignore the fourth band onwards. The error was minimized by simulated annealing, and then further optimized with the Nelder-Mead method. The parameters fitted to \cite{zhang2023polarization} are given in Table I. At both of the twist angles $\theta=3.89^\circ$ and $2.14^\circ$, the first three Chern numbers of the fitted continuum model are consistent with those of the DFT calculations \cite{zhang2023polarization}.

\begin{figure}[t!]
\centering
\includegraphics[width=0.99\linewidth]{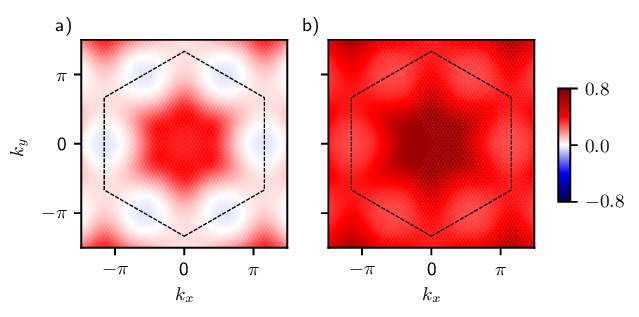}
\caption{Quantum geometry of the $\theta=2.1^\circ$ second moir\'{e} band at charge neutrality ($\nu_h=0$). (a) Berry curvature and (b) trace of the Fubini-Study metric. The black dashed line represents the moiré BZ with $a_M=1$.}
\label{fig:S_NI_2MB_QG}
\end{figure}
\paragraph{Quantum Geometry of the Second Moir\'{e} Band at Charge Neutrality}
In [Fig.\ref{fig:S_NI_2MB_QG}], we show the Berry curvature and trace of Fubini-Study metric of the second moiré band at charge neutrality. Compared with [Fig.1(c-d)] in the main text, we observe that both are less uniform, with the Berry curvature even undergoing a sign change near the $M$ points. {The quantum weight is $\frac{1}{2\pi}\int_\mathrm{BZ}\diff^2\mathbf{k}\:\trace\eta(\mathbf{k}) \approx 3.28$, which is further away from the ideal value of $3$ than the value of the SCHF-corrected band at $\nu_h=2$.}

\section{Self-Consistent Hartree-Fock Method}
Here we provide details on the self-consistent Hartree-Fock method used throughout the main text. All SCHF calculations were performed enforcing $\mathrm{U}_\uparrow(1)\times\mathrm{U}_\downarrow(1)$ charge symmetry, discrete translational symmetry, and time-reversal symmetry by the optimal damping algorithm \cite{cances2000can}. {We restrict our SCHF calculations at $\nu_h=2$ and $\nu_h=4$ to time-reversal symmetric integer spin Hall states observed in experiments \cite{kang2024observation,kang2024observation2}, while we intriguingly find that the valley-polarized states have lower energy within the SCHF. } For all SCHF calculations, we consider the mixing of the five highest moir\'{e} valence bands in each valley. The SCHF bands depicted in [Fig.1(b-d)] of the main text were performed on a $N_\text{HF}=24\times 24$ unitcell system. For the ED and pseudopotential calculations, we first calculated the SCHF-corrected bands on a more dense grid and took the subgrid, typically with $N_\text{HF}\geq 300$ sites. We have checked that the different choices of unitcells in calculation of SCHF and the subgrid does not significantly affect the bandwidth and quantum geometry.


{To obtain the band structures away from charge neutrality, we perform SCHF calculation over the continuum model fitted to the DFT results at charge neutrality instead of directly performing the DFT calculation at the desired integer hole filling. There are a few technical difficulties to conduct the DFT studies directly on the moir\'{e} flatband systems away from charge neutrality. Generally, DFT calculations with local or semi-local approximations for the exchange-correlation energy functional suffer from self-interaction error (also known as static correlation error or delocalization error) \cite{cohen2008insights}. Basically, the electronic structures of charged system with flat bands like the present case cannot be computed within DFT reliably. One may consider using further sophisticated methods to correct these errors such as hybrid functionals or the random phase approximation \cite{cohen2008insights}. However, the computational cost of doing so is prohibitively large due to the very large number of atoms in moiré unit cells. Moreover, to simulate doped layered systems using DFT, we need a further sophisticated method that also requires heavy computations \cite{brumme2014electrochemical,lee2021gate}, which makes the usage of the DFT more formidable.}



\section{Exact Diagonalization}
Here we provide implementation details such as finite crystal momentum grids, and additional results such as valley polarization and spectral flow.

\paragraph{Finite Crystal Momentum Grids.} We typically use $N=L_1\times L_2$ grids for the crystal momentum indexed by
\begin{equation}
\mathbf{k}_{k_1+L_1k_2}=\frac{k_1}{L_1}\mathbf{q}_1^M+\frac{k_2}{L_2}\mathbf{q}_2^M,
\label{eq:finite_crystal_momentum_grids}
\end{equation}
for $k_i\in\{0,1,\cdots,L_i-1\}$ for $i=1,2$, and a combined index $k_1+L_1k_2\in\{0,1,\cdots,N-1\}$. We additionally consider a $N=26$ grid,
\begin{equation}
\mathbf{k}_{k_1+2k_2}=\left\{\frac{k_1}{2}-\frac{2k_2}{13}\right\}\mathbf{q}_1^M+\frac{k_2}{13}\mathbf{q}_2^M,
\label{eq:finite_crystal_momentum_grid_26}
\end{equation}
for $k_1\in\{0,1\}$ and $k_2\in\{0,1,\cdots,12\}$, a combined index $k_1+2k_2\in\{0,1,\cdots,25\}$, where $\{x\}=x-\lfloor x\rfloor$ is the fractional part of $x$.

\begin{figure}[t!]
\centering
\includegraphics[width=0.99\linewidth]{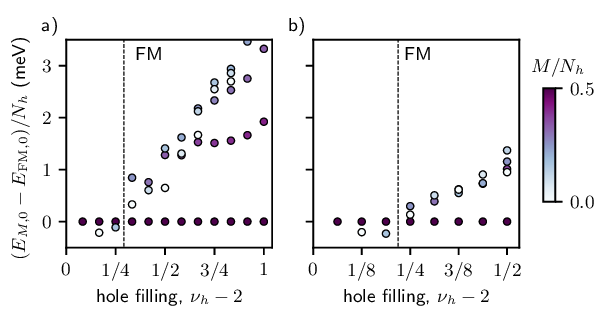}
\caption{Lowest energy of each magnetization sector relative to the FM sector in the $\theta=2.1^\circ$ second moir\'{e} band for $\epsilon_r=5$ as a function of hole filling, $\nu_h=N_h/N_\text{ED}$. Computed for (a) $N_\text{ED}=3\times 4$ and (b) $N_\text{ED}=4\times 4$ unit cells. The dashed line indicates the transition point where the ground state changes to the FM sector.}
\label{fig:S_2MB_magnetization_filling}
\end{figure}

\begin{figure}[b!]
\centering
\includegraphics[width=0.99\linewidth]{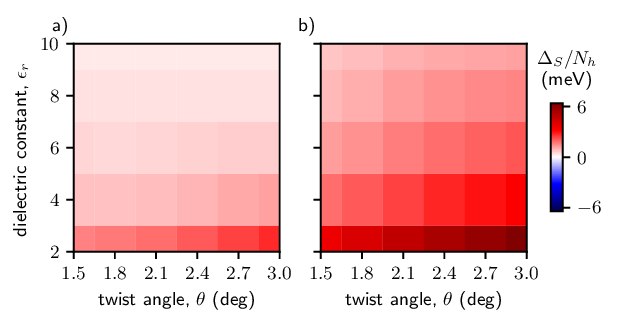}
\caption{Spin gap Eq.\eqref{eq:S_spin_gap} per hole between the FM sector and other magnetization sectors at (a) $\nu_h=2+1/2$ and (b) $\nu_h=3$ for $N_\text{ED}=3\times 4$ unitcells.}
\label{fig:S_2MB_magnetization_angle}
\end{figure}

\paragraph{Valley Polarization for $2<\nu_h\leq 3$.} Here we report exact diagonalization results about the valley polarization/ferromagnetism (FM) of the ground state for hole fillings $2<\nu_h\leq 3$, with particular emphasis on $\nu_h=2+1/2$ and $\nu_h=3$. All the calculations are performed on the SCHF-corrected second moir\'{e} bands at filling $\nu_h=2$, keeping both valleys. We denote $E_{M,0}$ as the lowest energy within the magnetization sector of magnetization $M$. The magnetization $M$ is defined as
\begin{equation}
M=\sum_iS_i^z.
\end{equation}
Because of the $\mathrm{U}_\uparrow(1)\times\mathrm{U}_\downarrow(1)$ symmetry of the continuum model and interaction, the magnetization is a conserved quantity.

In [Fig.\ref{fig:S_2MB_magnetization_filling}(a)], we show $E_{M,0}$ relative to the FM sector, $E_{\text{FM},0} = E_{M_{\text{FM}},0}$, for $\theta=2.1^\circ$ and $\epsilon_r=5$ as a function of filling factor $\nu_h=N_h/N_\text{ED}$ for $N_\text{ED}=3\times 4$ unitcells. An equivalent plot for $N_\text{ED}=4\times 4$ is shown in [Fig.\ref{fig:S_2MB_magnetization_filling}(b)], restricted to $\nu_h\leq 2+1/2$. We observe that the FM sector is energetically preferred at fillings $\nu_h=2+1/2$ and $\nu_h=3$.

Extending the parameter space to $\theta\in[1.5^\circ,3.0^\circ]$ and $\epsilon_r\in[2,10]$, while holding the other continuum model parameters fixed, we find that the ground state at $\nu_h=2+1/2$ and $\nu_h=3$ on a system with $N_\text{ED}=3\times 4$ unitcells is always fully-polarized FM. In [Fig.\ref{fig:S_2MB_magnetization_angle}], we plot the \textit{spin gap}, the lowest energy among other magnetization sectors relative to the ground state energy (always in the FM sector),
\begin{equation}
\Delta_S=\min_{M\::\:|M|\neq M_\text{FM}}\left\{E_{M,0}\right\}-E_{\text{FM},0}.
\label{eq:S_spin_gap}
\end{equation}

\begin{figure}[t!]
\centering
\includegraphics[width=0.99\linewidth]{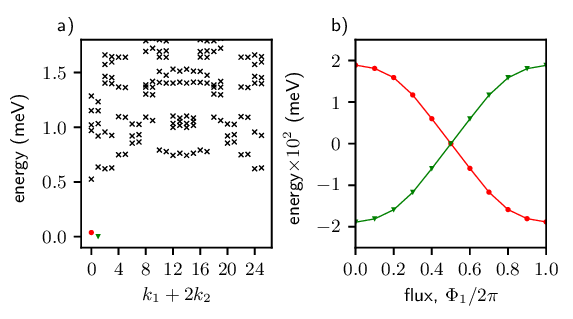}
\caption{{(a) Energy spectrum on a torus of $26$ unitcells. The two-fold ground states are colored as red and green. (b) Spectral flow of the ground states under flux insertion.}}
\label{fig:S_Pfaffian_N_26}
\end{figure}

{\paragraph{Energy Spectrum on system size $N=26$.} In [Fig.\ref{fig:S_Pfaffian_N_26}(a)] we show the energy spectrum for $N=26$, which clearly exhibits the two-fold ground state degeneracy of the pfaffian state \cite{read2000paired,oshikawa2007topological}.}

\begin{figure}[t!]
\centering
\includegraphics[width=0.99\linewidth]{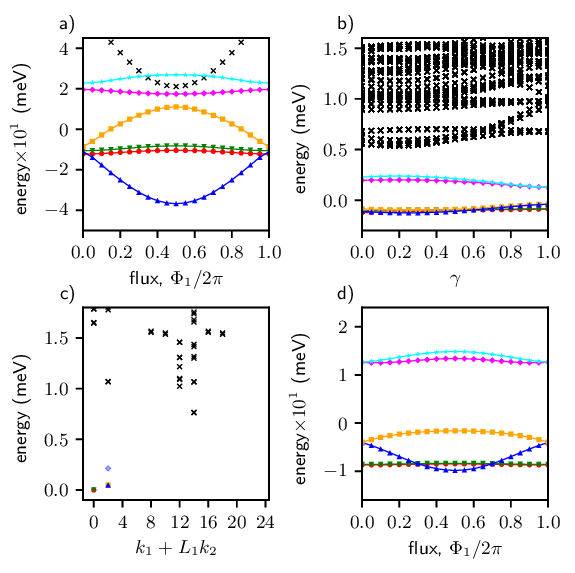}
\caption{{(a) Spectral flow of the ground states using the continuum model parameters in [Table 1] (of the main text) augmented with the SCHF-correction. This corresponds to the spectral flow of [Fig.2(a)] ($\Phi_1=0$) of the main text. (b) Evolution of the energy spectrum under linear interpolation of the continuum model parameters from the DFT-fitted parameters to Eq.(\ref{eq:modified_parameters}). (c) Energy spectrum and (d) spectral flow using Eq.(\ref{eq:modified_parameters}). All calculations are performed on a torus of $4\times 6$ unitcells, exhibiting six-fold ground state degeneracy. The ground states are shown with colors and connected by straight lines, while the excited states are shown with black 'x' markers.}}
\label{fig:S_Pfaffian_N_4_6_add}
\end{figure}

{\paragraph{Spectral Flow.} In [Fig.\ref{fig:S_Pfaffian_N_4_6_add}(a)] and [Fig.\ref{fig:S_Pfaffian_N_26}(b)], we show the spectral flow under flux insertion of the half-filled SCHF-corrected second moir\'{e} band at $\theta=2.1^\circ$ for $4\times 6$ and $26$ unitcells. We use $k_1\rightarrow k_1+\Phi_1/2\pi$ in Eq.(\ref{eq:finite_crystal_momentum_grids}) and Eq.(\ref{eq:finite_crystal_momentum_grid_26}). Consistent with the pfaffian state, we observe that the six and two approximately degenerate ground states for $N=4\times 6$ and $26$ evolve into each another under the flux insertion.

In [Fig.\ref{fig:S_Pfaffian_N_4_6_add}(a)], we observe that the flows of the ground states cross with the excited states while we can distinguish those of the ground states and excited states. To better resolve the $4\times 6$ spectral flow, we consider the adiabatically-connected, alternative model parameter and perform again the spectral flow calculation 
\begin{gather}
v_1=27.18\:\mathrm{meV},\:\psi=-58.29^\circ,\:v_2=-6.56\:\mathrm{meV},\nonumber\\
\gamma_1=-12.10\:\mathrm{meV},\:\gamma_2=18.73\:\mathrm{meV}.
\label{eq:modified_parameters}
\end{gather}
In [Fig.\ref{fig:S_Pfaffian_N_4_6_add}(b)], we show the evolution of the energy spectrum when the continuum model parameters interpolate linearly from the DFT-fitted parameters [Table.1] to the modified parameters of Eq.(\ref{eq:modified_parameters}). We find that the ground state evolves continuously between the two parameter sets, establishing the adiabatic connectivity between the two parameters. In [Fig.\ref{fig:S_Pfaffian_N_4_6_add}(c,d)], we show that with the modified continuum model parameters, the half-filled second moir\'{e} band exhibits six-fold ground state degeneracy and also the spectral flow of the six-fold degenerate ground states without crossing with excited states.}

{\section{On Similarity Metric Eq.(2)}
Here we provide additional details on the pseudopotential similarity metric. Specifically, we discuss general guidelines in the choice of the weights $w_\ell$ in Eq.(2) of the main text. There are two main considerations in setting the weights: (i) the first few $w_\ell$ should have a dominant contribution, and (ii) $w_\ell\rightarrow 0$ for large enough $\ell$. 

First, let us explain (i). It is well known that the first few pseudopotentials have the most significant influence on the nature of the emerging ground states. For example, see \cite{liu2013fractional,wang2015fermionic}. Notably, the Haldane pseudopotentials form monotonically decreasing sequences, so that the emerging ground states will favor to satisfy the first few Haldane pseudopotentials (when the ground state cannot satisfy all the pseudopotential components). That is, the first few pseudopotential components determine the nature of the emerging ground states. Also, the differences between the lowest, 1st, and 2nd LLs are most pronounced at the first few pseudopotentials, with higher LLs showing a slower drop-off of the first few terms, a feature which is suspected to play a significant role in stabilizing the Pfaffian state in 1LL \cite{peterson2008orbital,zaletel2013topological}. Hence, it is important to highlight the first few pseudopotentials. 

Second, on (ii), a cutoff in $\ell$ for $w_\ell$ is necessary due to practical implementation. The typical radius of the two-particle wavefunction with angular momentum $\ell$ is $r\approx\sqrt{2l_B\ell}$, where $l_B$ is the magnetic lengthscale \cite{tong2016lectures}. For very large $\ell$ where $r$ becomes comparable to the system size, the boundary effects will bias the values of the pseudopotentials, and may well deviate from those of the thermodynamic limit. Hence, to obtain the metric which is relatively free of this finite-size problem, it is safer to set $w_\ell\rightarrow 0$ for large $\ell$.

\begin{figure}[t!]
\centering
\includegraphics[width=0.99\linewidth]{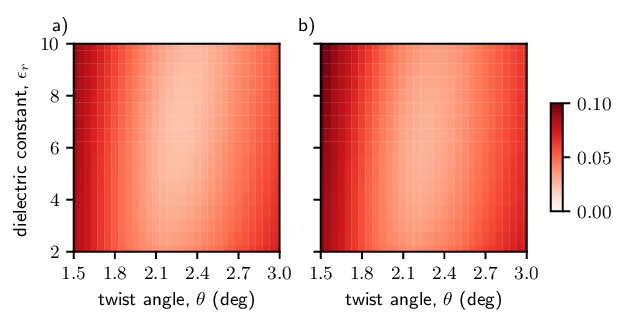}
\caption{{1LL-ness metric $S$ for the second moir\'{e} band, using the weights (a) Eq.(\ref{eq:similarity_weights_1}) and (b) Eq.(\ref{eq:similarity_weights_2}).}}
\label{fig:S_1LL_similarity_alts}
\end{figure}

Let us finally demonstrate the relative insensitivity of the similarity metric $S$ (of Eq.(2) in the main text) on the detailed choice of $w_\ell$ for our purpose, as long as the two guiding principles outlined above are met. In the main text, we have specifically used the step function, $w_\ell=1$ for $\ell\leq 9$ and $w_\ell=0$ for $\ell>9$. In [Fig.\ref{fig:S_1LL_similarity_alts}], we show the 1LL-ness metric $S$ of Eq.(2) using the alternative choices of the weights 
\begin{equation}
w_\ell^{(\text{exp})}=\begin{dcases}
\frac{1}{2^{(\ell-1)/2}} & \ell\leq 19\\
0 & \ell>19
\end{dcases},
\label{eq:similarity_weights_1}
\end{equation}
and
\begin{equation}
w_\ell^{(\text{norm})}=\begin{dcases}
\exp\left(-\frac{\ell^2}{50}\right) & \ell\leq 19\\
0 & \ell>19
\end{dcases}.
\label{eq:similarity_weights_2}
\end{equation}
Apparently, we see no significant discrepancies from [Fig.4(b)] of the main text.} 

\begin{figure}[t!]
\centering
\includegraphics[width=0.99\linewidth]{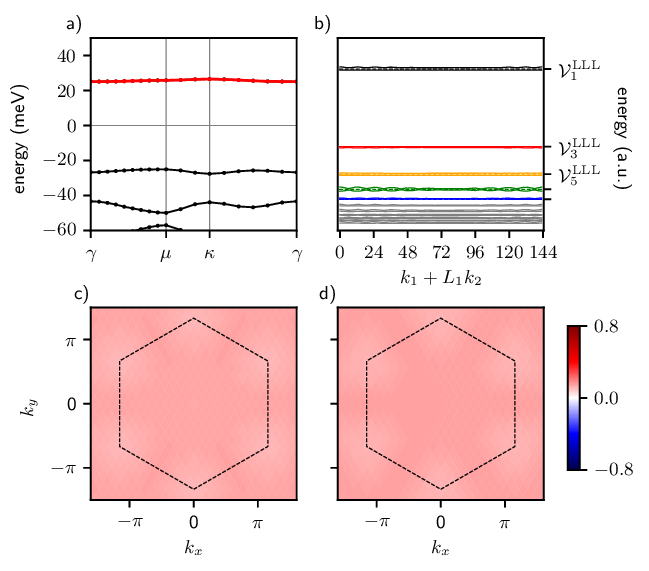}
\caption{{(a) SCHF-corrected moir\'{e} band structure at $\nu_h=1$ with $\epsilon_r=5$ and $\theta=2.1^\circ$. (b) Effective Haldane pseudopotentials, (c) Berry curvature, and (d) trace of the Fubini-Study metric of the SCHF-corrected first moir\'{e} band. The black dashed line represents the moir\'{e} BZ with $a_M=1$.}}
\label{fig:S_1MB_QG}
\end{figure}

\begin{table}[b!]
\begin{tabular}{c|ccccc|c}
\hline
angle & BW (meV) & $\sigma\left[\Omega(\mathbf{k})\right]$ & $\sigma\left[\trace\eta(\mathbf{k})\right]$ & QW & $S$ & Ref.\\
\hline\hline
$3.7^\circ$ & 11.2 & 0.05 & 0.04 & 1.09 & 0.115 & \cite{wang2024fractional}\\
-           &  8.8 & 0.04 & 0.04 & 1.14 & 0.106 & \cite{wang2023topology}\\
-           & 17.6 & 0.07 & 0.06 & 1.11 & 0.165 & \cite{zhang2023polarization}\\
\hline
$2.1^\circ$ &  1.4 & 0.01 & 0.01 & 1.02 & 0.016 & \cite{zhang2023polarization}\\
\hline
\end{tabular}
\caption{{Comparison of the $\nu_h=1$ SCHF-corrected first moir\'{e} bands at $\theta=2.1^\circ$ and those at $\theta=3.7^\circ$ and $\epsilon_r=5$. BW represents the bandwidth, $\sigma\left[\Omega(\mathbf{k})\right]$ and $\sigma\left[\trace\eta(\mathbf{k})\right]$ are the standard deviation of Berry curvature and quantum geometry, both calculated with the normalization $a_M=1$, QW is the quantum weight $\frac{1}{2\pi}\int_{\mathrm{BZ}}\diff^2\mathbf{k}\:\trace\eta(\mathbf{k})$, and $S$ is the LLL-ness of the projected interactions Eq.(2) of the main text. For $S$, we used $w_\ell=1$ for $\ell\leq 9$ and $w_\ell=0$ for $\ell>9$. Ref. indicates the literature from which we borrow the continuum model parameters.}}
\label{tab:LLL_similarity_angle}
\end{table}

{\section{First and Third Moiré Bands}
In the main text, we focused on the similarities of the second moir\'{e} band and the 1LL. Here we discuss the nature of the first and third moir\'{e} bands, which are in turn similar to LLL and 2LL.

\begin{figure}[b!]
\centering
\includegraphics[width=0.99\linewidth]{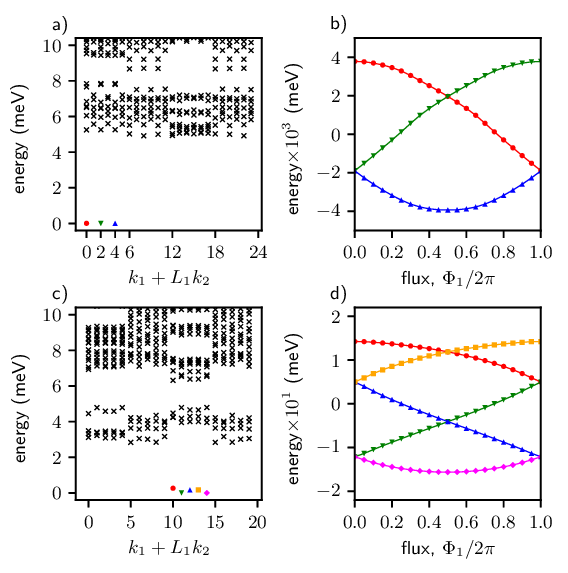}
\caption{{(a) Energy spectrum and (b) spectral flow under the flux insertion at $\nu_h=2/3$ filling on a torus of $6\times 4$ unitcells. (c) Energy spectrum and (d) spectral flow under the flux insertion at $\nu_h=3/5$ filling on a torus of $5\times 4$ unitcells. The ground states are shown with colors and connected by straight lines, while the excited states are shown with black 'x' markers.}}
\label{fig:S_1MB_FCIs}
\end{figure}

\paragraph{First Moiré Band.} Similar to the previous literature on $\theta=3.7^\circ$ \cite{devakul2021magic,xu2024maximally,dong2023composite,reddy2023toward,morales2024magic,yu2023fractional,jia2023moire,crepel2023chiral,wang2024fractional,abouelkomsan2024band}, we observe that the first moir\'{e} band is similar to the LLL. This is evidenced by the flat Berry curvature, trace of Fubini-Study metric, satisfication of the trace condition [Fig.\ref{fig:S_1MB_QG}], and the stabilization of $\nu_h=2/3$ and $3/5$ FQAH states at the corresponding fillings of the $\nu_h=1$ SCHF-corrected $\theta=2.1^\circ$ first moir\'{e} band [Fig.\ref{fig:S_1MB_FCIs}]. 

Remarkably, the LLL-ness of the first moiré band in $\theta=2^\circ$ is \textit{much better} than that in $\theta=3.7^\circ$. In [Table.\ref{tab:LLL_similarity_angle}], we compare the bandwidth, Berry curvature flatness, trace of Fubini-Study metric flatness, quantum weight, and the LLL-ness metric of the projected interactions of $\nu_h=1$ SCHF-corrected first moir\'{e} band at $\theta=3.7^\circ$ and $2.1^\circ$ using the parameters of [Table.1]. The LLL-ness of the first moiré band also manifests as the emergence of the LLL-like FQAH states at $\nu_h=2/3$ and $3/5$, whose ED spectrum can be found in [Fig.\ref{fig:S_1MB_FCIs}]. We clearly observe the three-fold and five-fold ground state degeneracies, each at $\nu_h=2/3$ and $\nu_h=3/5$, with the characteristic spectral flows. }

\begin{figure}[t!]
\centering
\includegraphics[width=0.99\linewidth]{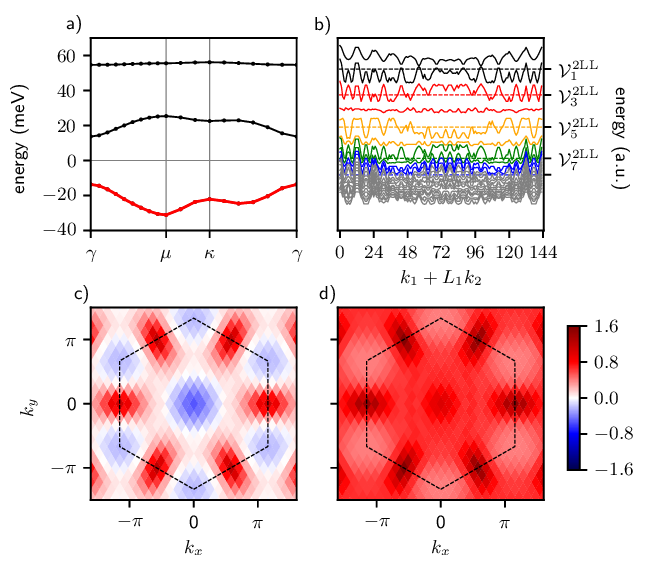}
\caption{{(a) SCHF-corrected moir\'{e} band structure at $\nu_h=4$ with $\epsilon_r=5$ and $\theta=2.1^\circ$. (b) Effective Haldane pseudopotentials, (c) Berry curvature, and (d) trace of the Fubini-Study metric of the SCHF-corrected third moir\'{e} band. The black dashed line represents the moir\'{e} BZ with $a_M=1$.}}
\label{fig:S_3MB_QG}
\end{figure}


\paragraph{Third Moiré Band} Here we report the quantum geometry and effective Haldane pseudopotentials of the SCHF-corrected third moir\'{e} band for $\theta=2.1^\circ$, $\epsilon_r=5.0$ at filling $\nu_h=4$. In [Fig.\ref{fig:S_3MB_QG}], we show the SCHF-corrected bands, and the effective Haldane pseudopotentials, Berry curvature, and trace of Fubini-Study metric of the third moir\'{e} band. We observe that the band is highly dispersive and the Berry curvature varies significantly across the moir\'{e} Brillouin zone. The quantum weight is $\frac{1}{2\pi}\int_\mathrm{BZ}\diff^2\mathbf{k}\:\trace\eta(\mathbf{k}) \approx 5.06$, close to the value 5, which is expected for a 2LL-like band \cite{ozawa2021relations,fujimoto2024higher}. The large band dispersion and non-ideal quantum geometry is reflected in the strong momentum dependence of the effective Haldane pseudopotentials, but the overall behavior does remain close to the exact 2LL pseudopotentials.


\section{Phase Diagram}
Here we provide details of how the phase diagram (Figure 4) was drawn. Let us denote $E_{K,a}$ ($a=1,2,3,\cdots$) as the energy level of each total crystal momentum sector, indexed by $K$, in ascending order. Note that $K=k_1+L_1k_2$ for the total crystal momentum $\mathbf{K}=k_1\mathbf{q}_1^M+k_2\mathbf{q}_2^M$. This parameterization is also used for [Fig.2] and [Fig.3] of the main text. We identify the non-Abelian FQAH in ED when the energy spectrum satisfies
\begin{equation}
\max\left(E_{0,2},E_{2,4}\right)\geq\min_{K\neq 0,2}\left\{E_{K,1}\right\}.
\end{equation}
We note that this is only a crude yet qualitative estimate of the FQAH region. First, as with all other ED calculations, because we are looking only at small system sizes, it is possible that a different ground state may appear in the thermodynamic limit. Second, because it is possible for certain anyon sectors to be energetically less-preferred to excited states while the ground state remains a FQAH state at finite sizes, this metric may systematically underestimate the FQAH region.

\bibliographystyle{apsrev4-1}
\bibliography{refs.bib}